\documentclass[11pt]{article}

\usepackage{graphicx}                   
\usepackage{amssymb,subfigure}   
\usepackage{verbatim}
\usepackage{color}
\usepackage{xcolor}
\usepackage{amsmath, xparse}

\def\beq{\begin{equation}}
\def\eeq{\end{equation}}
\def\bea{\begin{eqnarray}}
\def\eea{\end{eqnarray}}

\newcommand{\ec}{\end{center}}
\newcommand{\bc}{\begin{center}}
\newcommand{\pa}{\partial}
\newcommand{\mb}{\mathbf}

\begin{document}

\begin{center}
{\Large{\bf Strain Fields and Critical Phenomena in Manganites II: Spin-Lattice-Energy Hamiltonians}} \\
\ \\
\ \\
by \\
Rohit Singh and Sanjay Puri \\
School of Physical Sciences, Jawaharlal Nehru University, New Delhi -- 110067, India.
\end{center}

\begin{abstract}
The dynamic critical behavior at the paramagnetic-antiferromagnetic (PM-AFM) transition in manganites has recently been studied experimentally [Niermann et al., Phys. Rev. Lett. {\bf 114}, 037204 (2015)]. We extend the Hamiltonian of Paper I by incorporating an energy field, and study the corresponding {\it Model C} of critical dynamics. We use the dynamic renormalization group (RG) approach and calculate the dynamic critical exponents $z$, $\nu z$ and the line-width exponent $\Delta$ to leading order in the small expansion parameters $\epsilon=4-d+2\sigma$ and $\epsilon'=4-d$. Here, $d$ is the space dimension and $\sigma$ is the long-range exponent. Using $\sigma$ as an adjustable parameter, the theory gives us a good match to the experimentally available static and dynamic critical exponents at the PM-AFM transition.
\end{abstract}

\newpage

\section{Introduction}
\label{s1}

The dynamical behavior of systems near the critical point ({\it critical dynamics}) has attracted intense research interest \cite{sm76,ao04,am05,gm06}. Analytical studies of critical dynamics rely on a coupled set of Langevin-type stochastic equations of motion for the slow variables, i.e., the order parameters and hydrodynamic modes associated with conservation laws \cite{hh77}. Several different models have been proposed depending on conservation laws, leading to diverse universality classes. These classes are characterized by the dynamic critical exponent $z$, which connects the divergence of the time-scale ($\tau$) and the correlation length ($\xi$) near the critical point as $\tau\sim \xi^z$.

Dynamic critical phenomena have been studied extensively via  renormalization group (RG) analysis \cite{wf72,kw72,hhm74,hhs74,hhm76}. These RG techniques were originally developed to understand phase diagrams arising from model Hamiltonians. The family of critical exponents associated with a critical fixed point (FP) define the universality class of a phase transition. In the basic models of critical dynamics, e.g., {\it Model A} and {\it Model B} \cite{hh77}, the dynamics of the order parameter is non-conserved and conserved, respectively. There are also more sophisticated  models, e.g., {\it Model C}, with coupling of a non-conserved order parameter to a conserved energy density. The order parameter relaxes with a rate $\Gamma_0$, and the conserved density has a diffusion rate $\mu_0$ \cite{hh77}. A dynamical RG analysis predicts three distinct regions of phase space corresponding to the stable FP $f^*$ for the ratio $\Gamma_0/\mu_0$: {\it Region I} for $f^*=0$ (corresponds to $n>4$, where $n$ is the number of components of the order parameter); {\it Region II} for $f^*=n/(2-n)<\infty$ (corresponds to $n<2$); and {\it Region III} for $f^*=\infty$ (corresponds to $2<n<4$). In Region II, the conserved density influences the dynamics of the order parameter so that both variables are characterized by the same $z$ which differs from that of {\it Model A} ($z=2$).

The dynamics of the diffusive field becomes very important for systems where there is slow conduction of heat near their transition points. This diffusive coupling plays an essential role in describing the critical dynamics of, e.g., uniaxial AFMs \cite{hh77}, long-wavelength fluctuations near the QCD critical point \cite{br00}, aging dynamics at criticality \cite{cg03}. Clearly, the value of $z$ for {\it Model C} depends upon the conservation laws, dimensionality $d$, and the number of the components of the order parameter $n$. However, all experimental systems are not characterized by this clean universal behavior \cite{ms74,sq11}. As discussed in Paper I of this exposition, the static critical behavior in manganites at the paramagnetic-ferromagnetic (PM-FM) transition shows non-universality with continuously varying exponents different from those for short-range models. In I, we have shown that a model with spin-strain coupling proves very useful in predicting the nontrivial universality classes in manganites.  In this paper (II), we will turn our attention to critical dynamics in manganites in the vicinity of the paramagnetic-antiferromagnetic (PM-AFM) critical point.

Experimentally, a continuous variation of critical exponents has been reported for quite some time in various AFM  multiferroic manganites \cite{kmm01,tyk05,rgp05,hhk06,plg07,cgs07,osp12}. These systems show ferroelectric ordering at very high temperatures ($T_c\sim$1000 K) and AFM ordering at low temperatures ($T_N\sim$100 K). As a result, magnetic properties can be altered by changing the electric field, and vice versa \cite{ems06,hcs97,cm07}. For example, $\rm MnWO_4$ exhibits ferroelectricity induced by helical magnetic ordering \cite{hhk06,tat06,aa06,sta08}. Some experimental works \cite{flf02,lph05,lpk08} provide evidence for coupling of the order parameters of these two distinct phases, which is also responsible for a spin-lattice coupling \cite{pmh07,fpm09}. In addition, Olega et al. \cite{osp12} observed slow heat conduction near the PM-AFM transition point. 

In this paper, we aim to explain the continuously varying static and dynamic universality classes of AFM manganites \cite{osp12,plg07,ngb15}. For instance,  in multiferroic hexagonal $\rm RMnO_3$,  where $\rm R=Tm, Yb, Ho, Y, Lu, Er$, different choices of R lead to varying critical behavior with different values of the specific heat exponent $\alpha$ \cite{osp12}. Among various  materials, $\rm MnWO_4$ is the one that exhibits ferroelectricity induced by helical magnetic ordering \cite{hhk06,tat06,aa06,sta08,fly11,ngb15}. Niermann et al. \cite{ngb15} studied critical slowing down near the PM-AFM phase transition in single-crystal $\rm MnWO_4$ using broadband dielectric spectroscopy. They reported $\nu z \simeq 1.3$, where $\nu$ is the correlation length exponent ($\xi \sim |T-T_c|^\nu$). For the same sample, ultrasound experiments \cite{fly11} yield the magnetization exponent $\beta \simeq 0.45$. These values for $\nu z$, $\beta$ in $\rm MnWO_4$, and $\alpha$ in $\rm RMnO_3$, demonstrate that these manganites belong to universality classes different from those of the short-range models of critical dynamics (e.g., {\it Model A}, {\it Model C}) \cite{sm76}.

There have been some theoretical attempts to understand the critical behavior at the PM-AFM transition. Kawamura \cite{hk92,hk98} has performed RG calculations to study the PM-AFM transition in triangular AFMs. His results show that PM-AFM transitions in manganites do not belong to the universality classes of triangular AFMs and require a separate treatment. Further, the Green's function technique \cite{waa11,bw12} has also been invoked to explain the critical scenario in these systems, but the results are not convincing in respect of varying universality classes. We have seen in I that a long-ranged (LR) spin-lattice Hamiltonian can capture the widely varying universality classes in manganites near the PM-FM transition. The above-mentioned experiments emphasize that a spin-lattice coupling with slow heat modes dominates these systems \cite{lph05,pmh07,coh12}. Thus, a natural extension of our approach in I to the present context is by incorporation of spin-energy interactions. With this background, we consider a {\it Model C}-type spin-lattice-energy Hamiltonian \cite{hh77}:
\bea
{\cal H}[\Phi, \psi, \varepsilon] &=& \int d^d x\left[ \frac{r_0}{2}\Phi^2(\mb x) + \frac{c_0}{2} |\nabla\Phi(\mb x)|^2 + \frac{1}{2}\int d^d x'\psi(\mb x) u(\mb x-\mb x')\psi(\mb x') \right. \nonumber \\
&& \left. +g_0 \psi(\mb x) \Phi^2(\mb x)+\frac{e_0^{-1}}{2}\varepsilon^2(\mb x)+\gamma_0 \varepsilon(\mb x) \Phi^2(\mb x) \right] .
\label{staticH}
\eea
In Eq.~(\ref{staticH}), $\Phi$ is an $n$-component magnetic order parameter (with components $\phi_i$), $\psi$ is the lattice strain, and $\varepsilon$ is the energy field. As in I, we take the strain-strain interactions to be LR, i.e., $u(\mb x-\mb x')=\kappa_0^{-1}/|\mb x-\mb x'|^{d+2\sigma}$, where $\kappa_0^{-1}$ is the coupling constant, and $\sigma$ is the exponent of the LR interaction. The parameters $c_0$, $r_0$ and $e_0^{-1}$ have their usual interpretation. The terms with coupling $g_0$ and $\gamma_0$ account for the spin-lattice and spin-energy coupling, respectively. We formulate Model C with the above Hamiltonian, and study it using RG to calculate static and dynamic critical exponents.

Before proceeding, we remark that there have been several earlier studies of LR generalizations of the standard $\Phi^4$ Hamiltonian. One class of studies has focused on Ginzburg-Landau Hamiltonians where the $\Phi^2$-term is generalized to an LR form \cite{fmn72,dtc15,apr14,msn15,mbs17}. These models exhibit many novel features, e.g., multi-critical universality classes \cite{dtc15}, crossover phenomena \cite{apr14}, and stripe phases \cite{msn15,mbs17}. Further, the effect of LR generalizations of $\Phi^4$ interactions near the PM-FM transition was studied by Goll and Kopietz \cite{gk18}, who observed a finite Fisher exponent in $d=3$. Lattice models with LR spin interactions have also been studied by various authors \cite{mb93,frm07,apr14,lzd16} and shown to exhibit non-universal behavior, e.g., continuously varying Ashkin-Teller-like universality classes \cite{lzd16}.

This paper is organized as follows. In Sec.~\ref{s2}, we use RG analysis to identify the nontrivial FP and obtain the static exponents. We calculate the exponents to the leading order of a double expansion in $\epsilon=4-d+2\sigma$ and $\epsilon'=4-d$. In Sec.~\ref{s3}, we write down Model C and use dynamic RG to calculate the dynamic exponents. We then compare $\nu z$, $\beta$ and $\alpha$ with available experimental estimates. Finally, we conclude this paper and this two-part exposition with a summary and discussion in Sec.~\ref{s4}.

\section{Static Renormalization Group and Critical Exponents}
\label{s2}

In this section, we perform a static RG analysis of the Hamiltonian in Eq.~(\ref{staticH}). As we are interested in the long wavelength fluctuations, the real-space Hamiltonian can be transformed to Fourier space by using the $d$-dimensional Fourier transform for fields $f(\mb x)$ :
\beq
f(\mb x)=\int \frac{d^{d}k}{(2\pi)^d} f(\mb k)e^{i\mb k\cdot\mb x},
\eeq  
where $f$ is either of $\phi_i$, $\psi$, $\varepsilon$ or $u$. This yields
\bea
{\cal H} &=& \sum_{i=1}^{n} \int_0^\Lambda \frac{d^d  k}{(2\pi)^{d}}\frac{r_0+c_0k^{2}}{2}|\phi_{i}(\mb k)|^{2} 
+ \frac{1}{2} \int_0^\Lambda \frac{d^d k}{(2\pi)^{d}} u(\mb k)|\psi(\mathbf k)|^{2} \nonumber \\
&& + \int_0^\Lambda \frac{d^d k}{(2\pi)^{d}}
\frac{{e_0}^{-1}}{2} |\varepsilon (\mathbf k)|^{2}
+ g_0\sum_{i=1}^{n}\int_0^\Lambda \int_0^\Lambda \frac{d^{d} k_1}{(2\pi)^{d}}\frac{d^{d} k_{2}}{(2\pi)^{d}} \psi(\mb k_{1})\phi_{i}(\mb k_{2})\phi_{i}(-\mb k_1-\mb k_2) \nonumber \\
&& + \gamma_0 \sum_{i=1}^{n} \int_0^\Lambda \int_0^\Lambda \frac{d^{d} k_1}{(2\pi)^{d}}\frac{d^{d} k_{2}}{(2\pi)^{d}}\varepsilon(\mb k_{1})\phi_{i}(\mb k_{2})\phi_{i}(-\mb k_1-\mb k_2) .
\label{staticHF}
\eea
Here, the coupling function $u(\mb k)$ is
\beq
u(\mb k)=\kappa_0^{-1} k^{2\sigma} .
\label{LRC}
\eeq  
(In I, we had taken $u(\mb k)=\kappa_0^{-1} (k^2 + m^2)^\sigma$ and set $m \rightarrow 0$ later.)

\begin{figure}
\centering\includegraphics[width=0.7\textwidth,height=!]{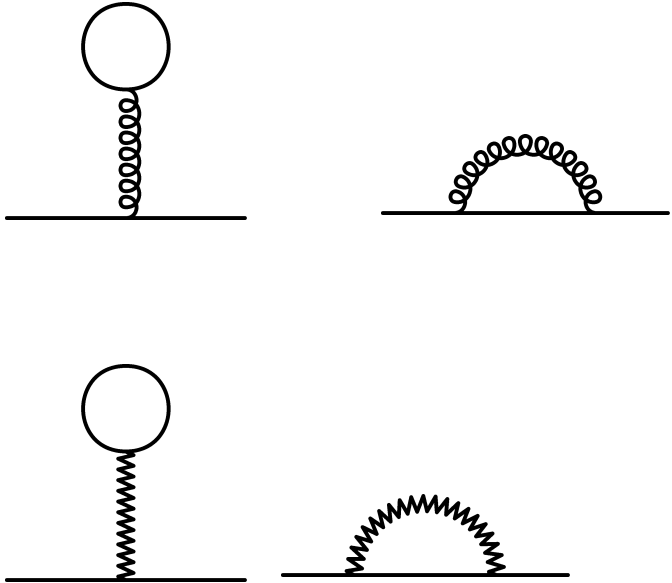}
\caption{Feynman diagrams giving self-energy corrections to $r_0$, $c_0$ at one-loop order. The external lines represent the $\Phi^{<}$ field. The internal straight, gluon and zigzag lines represent the correlation between $\Phi^>$, $\psi^>$ and $\varepsilon^>$ fields, respectively.}
\label{f1}
\end{figure}

\begin{figure}
\centering\includegraphics[width=0.65\textwidth,height=!]{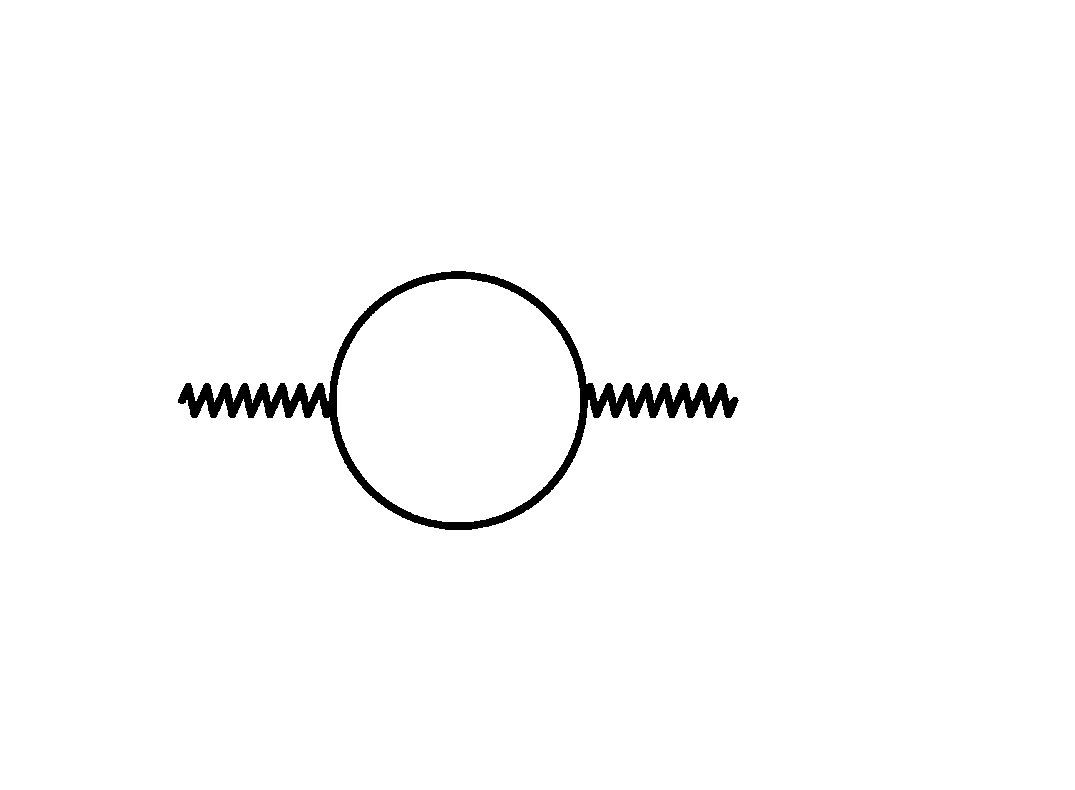}
\caption{Feynman diagram giving self-energy corrections  to $e_0^{-1}$ at one-loop order. The external lines represent the $\varepsilon^{<}$ field. The internal lines have the same meaning as in Fig.~\ref{f1}.}
\label{f2}
\end{figure}

\begin{figure}
\centering\includegraphics[width=0.6\textwidth,height=!]{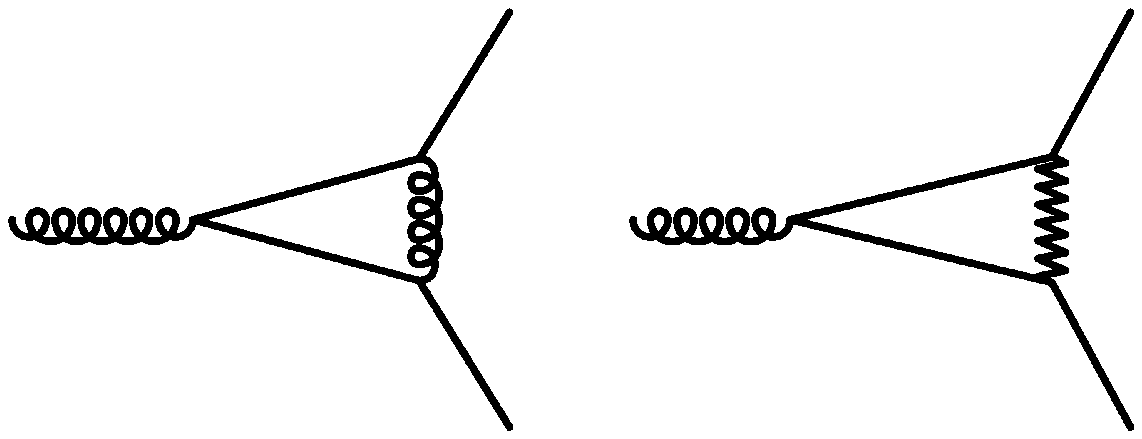}
\caption{Feynman diagrams for correction to $g_0$ at one-loop order. The notation is the same as in Fig.~\ref{f1}.}
\label{f3}
\end{figure}

\begin{figure}
\centering\includegraphics[width=0.6\textwidth,height=!]{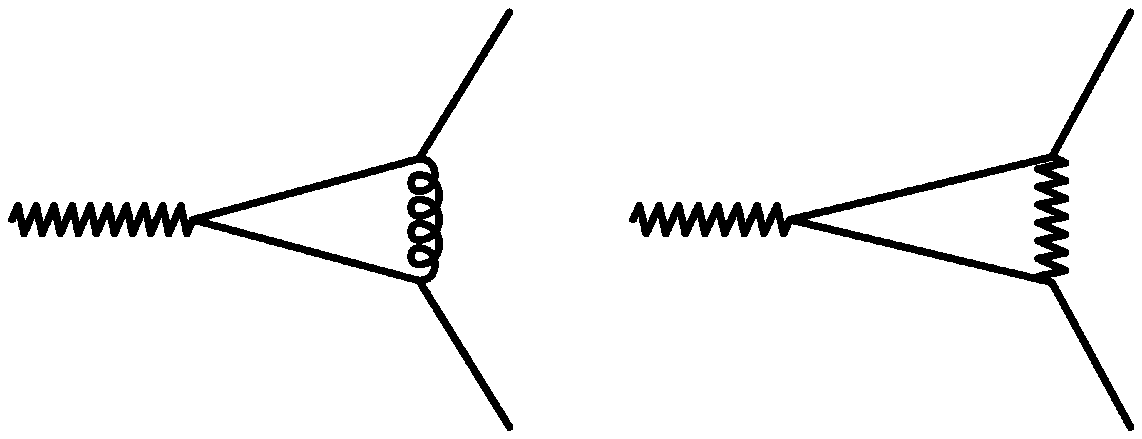}
\caption{Feynman diagrams for correction to $\gamma_0$ at one-loop order. The notation is the same as in Fig.~\ref{f1}.}
\label{f4}
\end{figure}

Using momentum shell decimation, we obtain the one-loop corrections contained in the amputated part of the diagrams given in Figs.~\ref{f1}-\ref{f4}. The elimination of modes $\phi_i^{>}(\mb k)$, $\psi^{>}(\mb k)$, and $\varepsilon^{>}(\mb k)$ lying in the momentum range $\Lambda/b \leqslant k\leqslant\Lambda$ yields the Hamiltonian  in terms of the remaining modes $\phi_i^{<}(\mb k)$, $\psi^{<}(\mb k)$, and $\varepsilon^{<}(\mb k)$ in the reduced range $0\leqslant k\leqslant \Lambda/b$. In this process, the bare parameters $r_0$, $c_0$, $\kappa_0^{-1}$, $e_0^{-1}$, $g_0$ and $\gamma_0$ acquire corrections that can be obtained by considering all the relevant Feynman diagrams in Figs.~ \ref{f1}-\ref{f4}. (We had presented details of the static RG procedure in I, and it would be redundant to repeat it here.) Incorporating these corrections, we obtain the following RG flow equations:
\bea
&& \frac{dr}{dl}=(2-\eta)r-\frac{4g^2\kappa S_d}{(2\pi)^d}\left(\frac{\Lambda^{d-2-2\sigma}}{c}-\frac{r}{c^2}\Lambda^{d-4-2\sigma}\right) \nonumber \\
&& \quad \quad \quad -\frac{(2n+4)\gamma^{2}e S_d}{(2\pi)^d}\left(\frac{\Lambda^{d-2}}{c}-\frac{r}{c^2}\Lambda^{d-4}\right), \label{flowr0} \\
&& \frac{dc}{dl}=-\eta c-\frac{4\sigma(2+2\sigma-d)g^2\kappa S_d}{d (2\pi)^d}\left(\frac{\Lambda^{d-4-2\sigma}}{c}\right), \label{flowc8} \\ 
&& \frac{d\kappa^{-1}}{dl}=(h-2\sigma)\kappa^{-1},\label{flowk} \\
&& \frac{de^{-1}}{dl}=(\alpha/\nu)e^{-1}-\frac{2n\gamma^{2}S_d}{(2\pi)^d}\left(\frac{\Lambda^{d-4}}{c^2}-\frac{r}{c^3}\Lambda^{d-6}\right),\label{flowC} \\
&& \frac{dg}{dl}=\left(\frac{4-d-2\eta+2\sigma}{2}\right)g+\frac{4g^3\kappa S_d}{(2\pi)^d}\left(\frac{\Lambda^{d-4-2\sigma}}{c^2}-\frac{2r}{c^3}\Lambda^{d-6-2\sigma}\right) \nonumber \\
&& \quad \quad \quad +\frac{4g\gamma^2 e S_d}{(2\pi)^d}\left(\frac{\Lambda^{d-4}}{c^2}-\frac{2r}{c^3}\Lambda^{d-6}\right), \label{flowgstaticg} \\
&& \frac{d\gamma}{dl}=\left(\frac{4-d-2\eta+\alpha/\nu}{2}\right)\gamma +\frac{4g^2 \kappa \gamma S_d}{(2\pi)^d}\left(\frac{\Lambda^{d-4-2\sigma}}{c^2}-\frac{2r}{c^3}\Lambda^{d-6-2\sigma}\right) \nonumber \\
&& \quad \quad \quad +\frac{4 \gamma^3 e S_d}{(2\pi)^d}\left(\frac{\Lambda^{d-4}}{c^2}-\frac{2r}{c^3}\Lambda^{d-6}\right),\label{flowgamma} 
\eea
where $b=e^{\delta l}$. In these equations, the exponents $\eta$ and $h$ are obtained from the two-point spin-spin correlation [$\sim r^{-(d-2+\eta)}$] and strain-strain correlation [$\sim r^{-(d-h)}$] functions at the transition point. Further, $S_d= 2\pi^{d/2}/\Gamma(d/2)$ is the surface area of a unit sphere in $d$ space dimensions; and $\alpha/\nu$ is the ratio of the specific heat exponent to the correlation length exponent. As in I, we define a new LR coupling $u=-g^2\kappa/2$. The flow equation for $u$ is 
\bea
\frac{du}{dl} &=& (4-d-2\eta+2\sigma)u -\frac{16u^2 S_d}{c^2(2\pi)^d}\left(\frac{\Lambda^{d-4-2\sigma}}{c^2}-\frac{2 r}{c^3}\Lambda^{d-6-2\sigma}\right) \nonumber \\
&& +\frac{8u\gamma^2 e S_d}{(2\pi)^d}\left(\frac{\Lambda^{d-4}}{c^2}-\frac{2r}{c^3}\Lambda^{d-6}\right) .
\label{flowlambda}
\eea 
 
We now redefine the dimensionless phase space parameters:
\bea
R &=& \frac{r}{\Lambda^2}, \nonumber \\
U &=& \frac{u S_d}{(2\pi)^d\Lambda^{4-d+2\sigma}}, \nonumber \\
F &=& \frac{\gamma^2e S_d}{(2\pi)^d\Lambda^{4-d}} ,
\label{flowRP8}
\eea
and  obtain the non-trivial FP. We notice that the redefined couplings $U$ and $F$ have scaling dimensions of $4+2\sigma$ and $4$, respectively. We find the upper critical dimensions from the marginality of these couplings as $d_c=4+2\sigma$ and $\bar{d_c} = 4$. We thus define the small expansion parameters
\bea
\epsilon &=& d_{c}-d=4-d+2\sigma , \nonumber \\
\epsilon' &=& \bar{d_{c}}-d=4-d .
\eea
The non-trivial FP is
\bea
\frac{R^*}{c} &=& \frac{\epsilon-2\epsilon'}{4} , \label{fp1} \\
\frac{U^*}{c^2} &=& \frac{\epsilon(n+4)-4\epsilon'}{16n}, \label{fp2} \\
\frac{F^*}{c^2} &=& \frac{\epsilon-\epsilon'}{2n}. \label{fp3}
\eea

We can linearize the flow equations for $R, U$ and $F$ around this FP to obtain the matrix equation:
\beq
\frac{d}{dl} \delta {\mb X} = {\mb M} \delta {\mb X} .
\eeq
Here, $\delta {\mb X}= {\mb X}- {\mb X}^*$ is the column vector $(\delta R, \delta U, \delta F)$, and ${\mb M}$ is a $3 \times 3$ matrix. The eigenvalues of ${\mb M}$ are 
\bea
\lambda_1 &=& 2+\frac{\epsilon}{2}-\epsilon' , \nonumber \\
\lambda_2 &=& \epsilon-\frac{2}{n}\left[\epsilon(n+4)-4\epsilon'\right]-\frac{4}{n}(\epsilon'-\epsilon) , \nonumber \\
\lambda_3 &=& \epsilon'-\frac{(2n+8)}{n}(\epsilon'-\epsilon)-\frac{1}{n}\left[\epsilon(n+4)-4\epsilon'\right] .
\eea
For critical properties, the conditions $\lambda_1>0$, $\lambda_2<0$ and $\lambda_3<0$ imply a stable FP. As in I, this fixes the range of $\sigma$-values. In $d=3$, the range of $\sigma$ varies with $n$:
\bea
-0.10 &<& \sigma<0~~\mbox{for}~~n=1, \nonumber \\
-0.17 &<& \sigma<0~~\mbox{for}~~n=2, \nonumber \\
-0.21 &<& \sigma<0~~\mbox{for}~~n=3.
\eea

The above results yield the ratio of $\alpha$ to $\nu$ in the leading order of $\epsilon$ and $\epsilon'$:
\beq
\frac{\alpha}{\nu}=\epsilon-\epsilon'+O(\epsilon^2,\epsilon\epsilon'+\epsilon'^2).
\label{alphanu}
\eeq
From the eigenvalue along the unstable direction $R$, we obtain $\nu$ as  
\beq
\nu=\frac{1}{2}+\frac{\epsilon'}{4}-\frac{\epsilon}{8}+O(\epsilon^2,\epsilon\epsilon',\epsilon'^2).\label{nu}
\eeq 
Using Eqs.~(\ref{alphanu})-(\ref{nu}) and well-known scaling relations \cite{sm76}, we obtain the static exponents:
\bea
\alpha &=& \frac{\epsilon-\epsilon'}{2}+O(\epsilon^2,\epsilon \epsilon',\epsilon'^2) , \label{alpha} \\ 
\beta &=& \frac{1}{2}-\frac{\epsilon}{8}+O(\epsilon^2,\epsilon\epsilon',\epsilon'^2) , \label{beta} \\
\gamma &=& 1+\frac{\epsilon'}{2}-\frac{\epsilon}{4}+O(\epsilon^2,\epsilon\epsilon',\epsilon'^2) , \label{gamma8} \\
\delta &=& 3+\epsilon'+O(\epsilon^2,\epsilon \epsilon',\epsilon'^2).\label{delta8}
\eea
These exponents will be compared with experimental results in Sec.~\ref{s3}, subsequent to the calculation of the dynamic exponent.

\section{Dynamic Renormalization Group}
\label{s3}

From Eq.~(\ref{staticH}), we see that the spin-lattice-energy Hamiltonian ${\cal H}$ is quadratic in the strain variable $\psi$ and energy field $\varepsilon$. As usual, the joint probability distribution function (jpd) is
\beq
P(\Phi,\psi,\varepsilon)=\frac{e^{-\beta {\cal H}}}{Z} , 
\eeq
where $\beta$ is the inverse temperature, and $Z$ is the appropriate partition function. This jpd for three variables can be reduced to a jpd for two variables by integrating the strain degrees of freedom. However, the effective Hamiltonian is now LR in the spin interactions with a redefined coupling constant. We study the critical dynamics for this effective Hamiltonian in the framework of $Model\,C$ \cite{hhm74}:
\bea
{\cal H}[\Phi, \varepsilon] &=& \int d^d x \left[\frac{r_0}{2} \Phi^2 (\mb x) + \frac{c_0}{2} |\nabla\Phi(\mb x)|^2 + \int d^d x'\Phi^2(\mb x) u(\mb x-\mb x') \Phi^2(\mb x') \right. \nonumber \\
&& \left. + \frac{e_0^{-1}}{2}\varepsilon^2(\mb x)+\gamma_0\Phi^2(\mb x)\varepsilon(\mb x)\right].
\label{hse}
\eea
In Eq.~(\ref{hse}), $u(\mb r)$ is LR in nature with a power law: $u(\mb r)=\lambda_0/r^{d+2\sigma}$.

To study the critical dynamics of ${\cal H}$ in Eq.~(\ref{hse}), we write the Langevin equations for the fields $\Phi (\mb x,t)$ and $\varepsilon (\mb x,t)$ as follows:
\bea
\frac{\pa}{\pa t}\Phi(\mb x,t) &=& -\Gamma_0\left(\frac{\delta {\cal H}}{\delta \Phi(\mb x,t)}\right)+ \mb \eta (\mb x,t),\label{Phi1} \\
\frac{\pa}{\pa t}\varepsilon(\mb x,t) &=& \mu_0\nabla^2\left(\frac{\delta {\cal H}}{\delta \varepsilon(\mb x,t)}\right)+\zeta(\mb x,t) . \label{e}
\eea
Here the constant $\Gamma_0^{-1}$ sets the time scale, and $\mu_0$ is the transport coefficient. The terms $\mb\eta(\mb x, t)$ and $\zeta(\mb x, t)$ represent Gaussian white noises with correlations 
\bea
\langle\eta_i(\mb x,t)\eta_j(\mb x',t')\rangle &=& 2\Gamma_0\delta^d(\mb x-\mb x')\delta(t-t')\delta_{ij}, \label{etan} \\
\langle\zeta(\mb x,t)\zeta(\mb x',t')\rangle &=& -2\mu_0\nabla^2\delta^d(\mb x-\mb x')\delta(t-t'). \label{zetan}
\eea
The angular brackets in Eqs.~(\ref{etan})-(\ref{zetan}) denote averages over the noise ensembles.

We introduce the $(d+1)$-dimensional Fourier transformation of the fields as 
\beq
f_i(\mb x,t)=\int_0^\Lambda \frac{d^d k}{(2\pi)^d}\int_{-\infty}^{\infty} \frac{d\omega}{2\pi}f_i(\mb k,\omega)e^{i(\mb k\cdot\mb x-\omega t)} ,
\eeq
where $f_i$ is either of $\phi_i$, $\epsilon$, $\eta_i$ or $\zeta$. Eqs.~(\ref{Phi1}) and (\ref{e}) can be written in Fourier space as
\bea
&& \left( -\frac{i\omega}{\Gamma_0}+r_0+c_0k^2\right)\phi_i(\mb k,\omega) = \frac{\eta_i(\mb k,\omega)}{\Gamma_0}-2\gamma_0\int\frac{d^d k_1 d\omega_1}{(2\pi)^{d+1}}\phi_i(\mb {k_1},\omega_1)\varepsilon(\mb k-\mb {k_1},\omega-\omega_1) \nonumber \\
&& -4 \sum_{j=1}^n \int \frac{d^d k_1 d\omega_1}{(2\pi)^{d+1}} \int\frac{d^d k_2 d\omega_2}{(2\pi)^{d+1}} u(\mb {k_1}-\mb k) \phi_i(\mb {k_1},\omega_1) \phi_j(\mb {k_2},\omega_2) \phi_j(\mb k-\mb {k_1}-\mb {k_2},\omega-\omega_1-\omega_2) , \label{ph} \nonumber \\ \\
&& \left( -\frac{i\omega}{\mu_0k^2}+e_0^{-1}\right)\varepsilon(\mb k,\omega) = \frac{\zeta(\mb k,\omega)}{\mu_0 k^2} -\gamma_0 \sum_{j=1}^n\int\frac{d^d k_1 d\omega_1}{(2\pi)^{d+1}}\phi_j(\mb k_1,\omega_1)\phi_j(\mb k-\mb{k_1},\omega-\omega_1) . \label{varepsilon}
\eea
The noise terms $\mb\eta(\mb k,\omega)$ and $\zeta(\mb k,\omega)$ have correlations in Fourier space as
\bea
\langle\eta_i(\mb k,\omega)\eta_j(\mb k',\omega')\rangle &=& 2\Gamma_0\delta_{ij} (2\pi)^{d+1}\delta^d(\mb k+\mb k')\delta(\omega+\omega'), \nonumber \\
\langle\zeta(\mb k,\omega)\zeta(\mb k',\omega')\rangle &=& 2\mu_0k^2 (2\pi)^{d+1}\delta^d(\mb k+\mb k')\delta(\omega+\omega'). \label{vef}
\eea

\begin{figure}
\centering\includegraphics[width=0.7\textwidth,height=!]{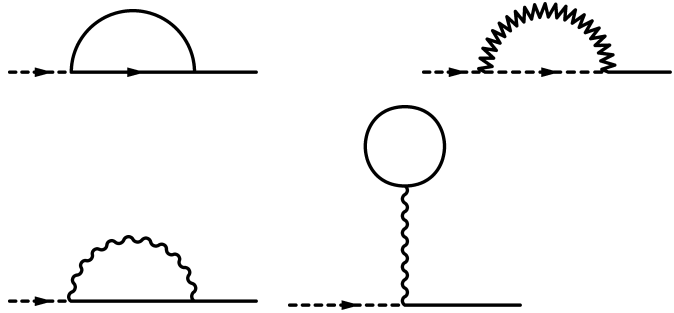}
\caption{Feynman diagrams for correction to the noise amplitude $\Gamma_0$ at one-loop order. The solid lines outside the loop represent the $\Phi$-field. Further, the dashed line with an arrow is the $G_0$ propagator; the solid line with an arrow is the $D_0$ propagator; and the wiggly line is the nonlocal coupling function. The straight and zigzag lines inside the loop represent the correlation between $\Phi^>$ and $\varepsilon^>$ fields, respectively.}
\label{f5}
\end{figure}

\begin{figure}
\centering\includegraphics[width=0.4\textwidth,height=!]{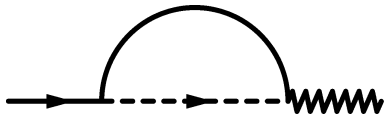}\caption{Feynman diagram for correction to the noise amplitude $\mu_0$ at one-loop order. The notation is the same as in Fig.~\ref{f5}.}
\label{f6}
\end{figure}

We need to find the relevant corrections to the dynamic parameters, namely, the noise amplitude $\Gamma_0$ and the transport coefficient $\mu_0$ at one-loop order. We define the bare propagators $G_0(\mb k,\omega)$ and $D_0(\mb k,\omega)$ as
\bea
G_0(\mb k,\omega) &=& \left(-\frac{i\omega}{\Gamma_0} + r_0 + c_0k^2\right)^{-1} , \\
D_0(\mb k,\omega) &=& \left(-\frac{i\omega}{\mu_0 k^2}+e_0^{-1}\right)^{-1} \label{propagators}.
\eea
In Eqs.~(\ref{ph})-(\ref{varepsilon}), we perform a dynamic RG analysis by eliminating the fast modes ($\Phi^>(\mb k,\omega)$, $\varepsilon^>(\mb k,\omega)$, etc.) lying in the band $\Lambda/b\leqslant k\leqslant \Lambda$. This yields equations for the slow modes ($\Phi^<(\mb k,\omega)$,  $\varepsilon^<(\mb k,\omega)$, etc.) belonging to $0\leqslant k\leqslant \Lambda/b$. In Figs.~\ref{f5} and \ref{f6}, we show the Feynman diagrams at one-loop order contributing to $\Gamma_0$ and $\mu_0$, respectively. The integration over fast modes yields
\bea
\phi_i^{<}(\mb k,\omega) &=& \frac{\eta_i(\mb k,\omega)}{\Gamma_0}G_0^{<}(\mb k,\omega) -4G_0^{<}(\mb k,\omega) \sum_{j=1}^n \int \frac{d^d k_1 d\omega_1}{(2\pi)^{d+1}} \int\frac{d^d k_2 d\omega_2}{(2\pi)^{d+1}} u(\mb {k_1}-\mb k) \times \nonumber \\
&& \phi_i^{<}(\mb {k_1},\omega_1) \phi_j^{<}(\mb {k_2},\omega_2) \phi_j^{<}(\mb k-\mb {k_1}-\mb {k_2},\omega-\omega_1-\omega_2) \nonumber \\ 
&& -2\gamma_0 G_0^{<}(\mb k,\omega)\int\frac{d^d {k_1}d\omega_1}{(2\pi)^{d+1}}\phi_i^{<}(\mb {k_1},\omega_1)\varepsilon^{<}(\mb k-\mb {k_1},\omega-\omega_1)+R_i(\mb k,\omega), \label{psi-se} \nonumber \\
\\
\varepsilon^{<}(\mb k,\omega) &=& \displaystyle\frac{\zeta(\mb k,\omega)}{\mu_0 k^2}D_0^{<}(\mb k,\omega) \nonumber \\
&& -\gamma_0D_0^{<}(\mb k,\omega) \sum_{i=1}^n \int\frac{d^d {k_1}d\omega_1}{(2\pi)^{d+1}}\phi_i^{<}(\mb{k_1},\omega_1)\phi_i^{<}(\mb k-\mb{k_1},\omega-\omega_1)+T(\mb k,\omega) . \label{e-se} \nonumber \\
\eea
In Eqs.~(\ref{psi-se})-(\ref{e-se}), the integration ranges for $\mb k_1$ and $\mb k_2$ are restricted to $0 \leq k_1,k_2 \leq \Lambda/b$. The quantities $R_i(\mb k,\omega)$ and $T(\mb k,\omega)$ give the corrections to the bare propagators $G_0(\mb k,\omega)$ and $D_0(\mb k,\omega)$, respectively. These can be written in terms of self-energies ($\Sigma_i, i=1,5$) as
\bea
R_i(\mb k,\omega) &=& -G_0^{<}(\mb k,\omega)\left[\Sigma_{1}(\mb k,\omega)+\Sigma_{2}(\mb k,\omega)+\Sigma_{3}(\mb k)+\Sigma_{4}(\mb k)\right]\phi_i^{<}(\mb k,\omega), \nonumber \\ \\
T(\mb k,\omega) &=& -D_0^{<}(\mb k,\omega)\Sigma_{5}(\mb k,\omega)\varepsilon^{<}(\mb k,\omega) .
\eea
The integral expressions for the self-energies are 
\bea
\Sigma_1(\mb k,\omega) &=& \frac{8\gamma_0^2}{\Gamma_0}\int\frac{d^d q d \Omega}{(2\pi)^{d+1}}D_{0}^{>}(\mb k-\mb q,\omega-\Omega)|G_0^{>}(\mb q,\Omega)|^2, \label{sigma1} \\
\Sigma_2(\mb k,\omega) &=& \frac{8\gamma_0^2}{\mu_0}\int\frac{d^d q d\Omega}{(2\pi)^{d+1}}q^{-2}G_0^{>}(\mb k-\mb q,\omega-\Omega)|D_0^{>}(\mb q,\Omega)|^2, \label{sigma2} \\
\Sigma_3(\mb k) &=& \frac{16}{\Gamma_0} \int\frac{d^d q d\Omega}{(2\pi)^{d+1}}u(\mb q-\mb k)|G_0^{>}(\mb q,\Omega)|^2, \label{sigma4} \\
\Sigma_4(\mb k) &=& \frac{8n}{\Gamma_0}u(\mb0)\int\frac{d^d q d\Omega}{(2\pi)^{d+1}}|G_0^{>}(\mb q,\Omega)|^2, \label{sigma3} \\
\Sigma_5(\mb k,\omega) &=& \frac{4n\gamma_0^{2}}{\Gamma_0}\int\frac{d^d q d\Omega}{(2\pi)^{d+1}}G_0^{>}(\mb k-\mb q,\omega-\Omega)|G_0^{>}(\mb q,\Omega)|^2 , \label{sigma5}
\eea
where the $\mb q$-integral ranges over $\Lambda/b \leq q \leq \Lambda$.
In the above self-energy integrals, we carry out the frequency convolutions by integration over the internal frequency $\Omega$ in the full range to obtain
\bea
\Sigma_1 &=& 4\gamma_0^2\mu_0 \int_{\Lambda/b}^\Lambda \frac{d^d q}{(2\pi)^{d}}\frac{(\mb k-\mb q)^2}{(c_0q^2+r_0)[-i\omega+\Gamma_0(c_0q^2+r_0)+\mu_0(\mb k-\mb q)^2/e_0]}, \\
\Sigma_2 &=& 4\gamma_0^2\Gamma_0e_0 \int_{\Lambda/b}^\Lambda \frac{d^d q}{(2\pi)^{d}}\frac{1}{[-i\omega+\Gamma_0(c_0q^2+r_0)+\mu_0(\mb k-\mb q)^2/e_0]} , \\
\Sigma_3 &=& 8 \int_{\Lambda/b}^\Lambda \frac{d^d q}{(2\pi)^{d}}u(\mb q-\mb k)\frac{1}{(c_0 q^2+r_0)}, \\
\Sigma_4 &=& 4nu(\mb 0) \int_{\Lambda/b}^\Lambda \frac{d^d q}{(2\pi)^{d}}\frac{1}{(c_0 q^2+r_0)}, \\
\Sigma_5 &=& 4n\gamma_0^2 \int_{\Lambda/b}^\Lambda \frac{d^d q}{(2\pi)^d}\frac{1}{(c_0q^2+r_0)\left\{-i\omega/\Gamma_0+(c_0q^2+r_0)+[c_0(\mb k-\mb q)^2+r_0]\right\}} . \nonumber \\
\eea

We see that $\Sigma_3$ and $\Sigma_4$ do not carry an $\omega$-dependence, and thus do not contribute to the noise amplitude $\Gamma_0$. The self-energy corrections $\Sigma_{i=1,2}(\mb k, \omega)$ modify the propagator $G_0$ and they  yield  corrections to $\Gamma_0^{-1}$ as
\beq
\Delta \Gamma^{-1}=\sum_{i=1}^2 \Delta\Gamma_i^{-1},
\eeq
where
\beq
\Delta\Gamma_{i}^{-1}= i\frac{\pa}{\pa\omega}\left[\Sigma_i(\mb k, \omega)\right]_{\mb k=0,\omega=0} .
\eeq  
Here,
\bea
\Delta\Gamma^{-1}_1 &=& 4\gamma_0^2\mu_0 \int_{\Lambda/b}^\Lambda \frac{d^d q}{(2\pi)^d}\frac{q^{2}}{(c_0q^2+r_0)[\Gamma_0(c_0q^2+r_0)+ \mu_0 q^2/e_0]^{2}}\label{gamma1}, \\
\Delta\Gamma^{-1}_2 &=& 4\gamma_0^2\Gamma_0e_0 \int_{\Lambda/b}^\Lambda \frac{d^d q}{(2\pi)^d}\frac{1}{\left[\Gamma_0(c_0q^2+r_0)+ \mu_0 q^2/e_0 \right]^{2}}.\label{gamma2}
\eea
Carrying out the integrations over the internal momentum in Eqs.~(\ref{gamma1})-(\ref{gamma2}), we obtain the total one-loop correction to the noise amplitude $\Gamma_0$ as
\beq
\Delta\Gamma^{-1}=4\frac{\gamma_0^2 e_0S_d}{c_0^2(2\pi)^d} \left[\Gamma_0\left(1+\frac{\varrho_0}{c_0}\right)\right]^{-1} \frac{(b^{4-d}-1)\Lambda ^{d-4}}{(4-d)} \label {delgamma}, 
\eeq
where $\varrho_0= \mu_0/(\Gamma_0 e_0)$. The self-energy correction $\Sigma_5$
modifies the propagator $D_0$, and it gives corrections to $\mu_0^{-1}$ as
\beq
\Delta\mu^{-1}=\lim_{k,\omega\rightarrow 0}i\frac{\pa}{\pa(\omega/k^2)}\Sigma_5(\mb k, \omega) . \label{lambda0}
\eeq
This quantity vanishes because $\Sigma_5(\mb k, \omega)$ does not involve the ratio $\omega/k^2$.

The RG transformation requires that the equations of motions and the correlations preserve their forms under the scale transformations. Thus, we write rescaled momentum and fields as 
\bea
\mb k'&=& b \mb k, \nonumber \\
\omega' &=& b^z\omega, \nonumber \\
\Phi'(\mb k', \omega') &=& b^{x}\Phi^{<}(\mb k, \omega), \nonumber \\
\epsilon'(\mb k', \omega') &=& b^{y}\epsilon^<(\mb k, \omega),
\eea
where
\bea
x &=& \frac{d-2+\eta}{2} , \nonumber \\
y &=& \frac{d-\alpha/\nu}{2} ,
\eea
and $z$ is the dynamic critical exponent. We thus arrive at the recursion relations for the noise amplitudes as   
\bea
(\Gamma^{-1})' &=& b^{2-z-\eta}(\Gamma_0^{-1}+\Delta\Gamma^{-1}) , \nonumber \\
(\mu^{-1})' &=& b^{d+2-2y-z}\mu_0^{-1}.
\eea
Using $\Delta\Gamma^{-1}$ from Eq.~(\ref{delgamma}), the recursion relations yield the RG flow equations:
\bea
\frac{d\Gamma^{-1}}{dl} &=& (2-z-\eta)\Gamma^{-1}+4\frac{\gamma^2 eS_d}{c^2(2\pi)^d}\left[\Gamma\left(1+\frac{\varrho}{c}\right)\right]^{-1}\Lambda^{d-4}, \label{Gamma} \\
\frac{d\mu^{-1}}{dl} &=& (d+2-2y-z)\mu^{-1}, \label{mu}
\eea
where $b=e^{\delta l}$. 

We also introduce $\varrho = \mu/(\Gamma e)$, which satisfies the flow equation
\beq
\frac{d\varrho}{dl}= \varrho \left[-\eta+4\frac{\gamma^2 eS_d}{c^2(2\pi)^d}\left\{\left(1+\frac{\varrho}{c}\right)\right\}^{-1}\Lambda^{d-4}-2n\frac{\gamma^2 eS_d}{c^2(2\pi)^d}\Lambda^{d-4}\right] . \label{flowg}
\eeq
Substituting the FP value of $\gamma^{2}e$ from Eq.~(\ref{fp3}) in Eq.~(\ref{flowg}), we obtain the non-trivial FP as 
\beq
\frac{\varrho^*}{c}=\frac{2}{n}-1.
\label{fp4}
\eeq
A linear stability analysis about this FP shows that it is stable for $n>2$ in $d=3$. (This should be contrasted to the original $\it Model\, C$ calculation \cite{hhm74}, where the FP is found to stable for $n<2$.) The dynamic critical exponent $z$ corresponding to this FP is
\beq
z=2+\epsilon-\epsilon'+O(\epsilon^2,\epsilon'\epsilon,\epsilon'^2) .
\label{z}
\eeq
As $\sigma$ is restricted to the range $-0.21<\sigma<0$ for $d=3$ and $n=3$ in Sec.~\ref{s2}, this means that $1.58 < z < 2$ for the same $d,n$.

Further, the linewidth exponent $\Delta$, given by the scaling relation $\Delta=\nu(z+2-d-\eta)$, is obtained as
\beq
\Delta=\frac{\epsilon}{2}+O(\epsilon^2,\epsilon'\epsilon,\epsilon'^2).
\label{w}
\eeq
We also calculate the exponent $\nu z$ (related to  the characteristic time scale as $\tau\propto |T-T_c|^{-z\nu}$) and obtain 
\beq
\nu z=1+\frac{\epsilon}{4}+ O(\epsilon^2,\epsilon'\epsilon,\epsilon'^2).
\label{nuz}
\eeq
Thus, we see that the values of $\Delta$ and $\nu z$ for $d=3,n=3$ lie in the range $0.29<\Delta<0.5$ and $1.145<\nu z<1.250$.

As in I, we compare our RG exponents with available experimental results. Unfortunately, there are not many measurements of the dynamical exponent. A comparison of our theory with experiments is shown in Table~\ref{comp}. In I, we had used the magnetization exponent $\beta$ to fix the vale of $\sigma$. Here, we use the specific heat exponent $\alpha$ from Eq.~(\ref{alpha}) to stress that our theory does not critically depend on how we fix $\sigma$. The other RG exponents are obtained from Eqs.~(\ref{alpha})-(\ref{delta8}) and Eqs.~(\ref{z})-(\ref{nuz}).

The experimental values for $\rm MnWO_4$, namely, $\nu z\approx 1.3$ \cite{ngb15} and $\beta=0.45$ \cite{fly11} are comparable to our RG values $\nu z=1.15$ and $\beta=0.43$ obtained for $\sigma=-0.20$, as shown in Table \ref{comp}. (In this case, we did not have an experimental value of $\alpha$, so $\sigma$ was fixed at the lower end of its acceptable range.) In addition, experiments yielded the specific heat exponent \cite{osp12} $\alpha=-0.09\pm0.01$ for ErMnO$_3$; $\alpha =-0.12\pm0.01$ for YMnO$_3$; $\alpha=-0.12\pm0.01$ for LuMnO$_3$; $\alpha =-0.14\pm0.01$ for HoMnO$_3$; $\alpha=-0.18\pm0.01$ for YbMnO$_3$; and $\alpha =-0.19\pm0.01$ for TmMnO$_3$. Also, an ultrasonic investigation \cite{plg07} of single-crystal YMnO$_3$ gives $\beta=0.42\pm 0.03$. These values of $\alpha$ and $\beta$ are reproducible from the present theory by appropriate choices of $\sigma$, as shown in Table~\ref{comp}. It will be no exaggeration to say that the present model is viable for exploring the static and dynamic critical behavior of AFM manganites near their transition temperatures. We urge experimentalists to undertake further studies of the PM-AFM critical point in manganites, with particular emphasis on measurements of the dynamical exponent.

\section{Summary and Discussion}
\label{s4}

We conclude this paper II and two-part exposition with a summary and discussion of our results. In II, we have explored the static and dynamic critical properties of antiferromagnetic (AFM) manganites near their PM-AFM transition. These materials exhibit ferroelectric ordering at high temperatures and AFM ordering at low temperatures. The order parameters of these two distinct phases couple, generating a strong spin-lattice coupling \cite{pmh07,fpm09}. In addition, thermal diffusivity data shows a slow heat conduction near the Neel temperature \cite{osp12}. In I, we have presented renormalization group (RG) calculations for a long range (LR) spin-lattice Hamiltonian. We have shown that this model can capture the non-universal critical behavior in manganites near their PM-FM transition point. In II, we focus on the effect of LR interactions on critical behavior in manganites near their PM-AFM transition. We formulate an LR spin-lattice-energy Hamiltonian [Eq.~(\ref{staticH})], and explore its equilibrium and non-equilibrium critical properties via RG analysis.

In this context, we first study static exponents in Sec.~\ref{s2}. We use these to study dynamic exponents in Sec.~\ref{s3}. Our static calculations were done on the Hamiltonian in Eq.~(\ref{staticH}). For the dynamic calculations, we studied Model C with an effective Hamiltonian [Eq.~(\ref{hse})] obtained by integrating out the strain degrees of freedom.

In Table~\ref{comp}, we have compared our one-loop RG results for static and dynamic critical exponents with available experimental results for AFM manganites. (Unfortunately, there is a paucity of experiments on dynamic critical phenomena in manganites.) For example, experiments have yielded the specific heat exponent $\alpha$ for different samples of RMnO$_3$  in the range $-0.20 \leq \alpha \leq -0.09$. Also, an ultrasonic investigation \cite{plg07} of YMnO$_3$ gave the magnetization exponent $\beta=0.42\pm 0.03$. We notice that these values of $\alpha$ and $\beta$ are quite different from those of short-range $d=3$ Heisenberg, Ising and XY models \cite{sm76}. It has been argued that a chiral universality class might be expected because of the triangular geometry. In this context, the RG results of Kawamura \cite{hk98} yielded $\alpha = 0.34$, $\beta = 0.25$ for the chiral XY model; and $\alpha=0.24$, $\beta=0.30$ for the chiral Heisenberg model. These numbers are also at variance with the experiments. However, the experimental exponents match well with our theoretical estimates in the allowed range of $\sigma$. For the dynamic critical exponent ($\nu z$) in MnWO$_4$, we find that the observed exponent ($\nu z \simeq 1.3$) is consistent with our RG value ($\nu z=1.15$) for $\sigma=-0.20$.

We would like to conclude by noting that the recent revival of interest in multiferroic materials is due to their multiple applications, e.g., magnetic recording read heads, photovoltaic multiferroic cells, etc. Clearly, the technological aspects of these systems are very exciting. In this two-paper exposition, we have shown via RG calculations that these systems also provide a playground where various interactions lead to rich and diverse physics near phase transitions. We hope that the approach presented here will motivate further research in such demanding compounds, where competing interactions lead to unconventional physics.

A possible extension of our work would be to study the far-from-equilibrium quench dynamics of such PM-FM and PM-AFM phase transitions. These symmetry-breaking transitions generate defects that could be topological in nature. In this context, there are two classes of interesting problems: (a) the dependence of the defect density on the quench protocol; and (b) the kinetics of defect annihilation subsequent to the quench, i.e., the domain growth or coarsening of the system \cite{pw09,dp04}. As we have shown explicitly here, the inclusion of strain fields yields widely varying universality classes in such systems. It is of great relevance to study the corresponding spectrum of coarsening problems and investigate the domain growth laws and evolution morphologies. Despite the useful implications of the current approach to understanding the critical dynamics of multiferroics, we believe our calculations also have important consequences for the far-from-equilibrium kinetics of phase transitions.

\subsubsection*{Acknowledgments}

R.S. is grateful to the University Grants Commission (UGC), India for financial support through a D.S. Kothari postdoctoral fellowship. We are grateful to the referees for their constructive comments and suggestions.

\newpage
\begin{table}[!]
{\caption{\small Experimental values of critical exponents for antiferromagnetic manganites, and their comparison with the  present theory. The value of $\alpha$ (wherever available) is used to fix the value of $\sigma$ in the allowed range for $d=3,n=3$. The other exponents $\beta$, $\nu$ and $\nu z$ correspond to this value of $\sigma$. Unavailable data is indicated by dashes.} \label{comp}}
\footnotesize
\begin{center}
\begin{tabular}{|c|c|c|c|c|c|}\hline
$\sigma$  & Theory/ & $\alpha$ & $\beta$ & $\nu$ & $\nu z$\\
& Experiments & & & & \\
\hline
$-0.20$ & Theory &   $-0.20$& $0.425$& $0.675$ & $1.150$\\
\cline{2-6}
& MnWO$_4$\cite{fly11,ngb15} & $-$ & $0.45$ & $-$ & $1.3$\\
\hline
$-0.19$ & Theory &   $-0.190$& $0.423$& $0.673$ & $1.155$\\
\cline{2-6}
&  TmMnO$_3$ \cite{osp12} & $-0.19\pm 0.01$ & $-$ & $-$ & $-$\\
\hline
$-0.18$ & Theory&  $-0.180$ & $0.420$& $0.670$ &  $1.160$\\
\cline{2-6}
&  YbMnO$_3$ \cite{osp12} & $-0.18\pm 0.01$ & $-$ & $-$ & $-$\\
\hline
$-0.14$ &Theory & $-0.140$ & $0.410$ & $0.660$ & $1.180$\\
\cline{2-6}
&  HoMnO$_3$ \cite{osp12}& $-0.14\pm 0.01$ & $-$ & $-$ & $-$ \\
\hline
$-0.12$ & Theory & $-0.120$& $0.405$ & $0.655$ & $1.190$ \\
\cline{2-6}
&YMnO$_3$ \cite{osp12,plg07}& $-0.12\pm 0.01$ & $0.42\pm 0.03$ & $-$ &$-$\\
\cline{2-6}
&  LuMnO$_3$\cite{osp12} &$-0.12\pm 0.01$ & $-$ & $-$ & $-$ \\
\hline
$-0.09$ & Theory&  $-0.090$ & $0.398$ & $0.648$ & $1.045$\\
\cline{2-6}
&  ErMnO$_3$ \cite{osp12} & $-0.09\pm 0.01$ & $-$ & $-$ & $-$\\
\hline
\end{tabular}
\end{center}
\end{table}
%\vspace{3cm}
\newpage

\end{document}